# Universal Design and Adaptive Interfaces as a Strategy for Induced Disabilities

**Aaron Steinfeld**
Carnegie Mellon University
5000 Forbes Ave.
Pittsburgh, PA 15213 USA
steinfeld@cmu.edu

**John Zimmerman**
Carnegie Mellon University
5000 Forbes Ave.
Pittsburgh, PA 15213 USA
johnz@cs.cmu.edu

**Anthony Tomasic**
Carnegie Mellon University
5000 Forbes Ave.
Pittsburgh, PA 15213 USA
anthony.tomasic@gmail.com

**ABSTRACT**
There is great promise in creating effective technology experiences during situationally-induced impairments and disabilities through the combination of universal design and adaptive interfaces. We believe this combination is a powerful approach for meeting the UX needs of people with disabilities, including those which are temporary in nature. Research in each of these areas, and the combination, illustrates this promise.

## INTRODUCTION

The concept of temporary disability due to unfavorable situations has a long history, especially in the context of noise-induced deafness [10] or environmental deafness (e.g., [13] p. 21). Anyone with good hearing invariably finds themselves unable to hear conversations at a noisy factory, social space, or musical event. In these contexts, systems and devices designed to support people with disabilities often have broader value to people without disabilities. Therefore, there are many relevant lessons and strategies from the accessible technology field that are relevant to the wider population.

Within human-computer interaction (HCI), this situation has been characterized by Sears et al. [12] as situationally-induced impairments and disabilities (SIIDs) and specifically where:

> "… both the environment in which an individual is working and the activities in which she is engaged can contribute to the existence of impairments that interfere with the use of computing technologies." (p.75)

One proven strategy from the accessible technology community that is relevant for SIIDs is universal design. Curb cuts are the traditional example given for universal design since they support sidewalk access to a wide range of users beyond wheelchair and scooter users (e.g., parents with strollers, people on bicycles, deliveries on carts, etc.). However, a better example in the context of information access is captioning on televisions in bars and other noisy environments.

In parallel, there has also been a recent emphasis on artificial intelligence (AI) and machine learning to provide tailored, context-aware assistance based on the user's location, plans, and previous behavior (e.g., location-based searches, driving directions to home, etc.). Experiences are becoming increasingly personalized as AI learns more about their users. Personalization through machine learning leads to new types of user interactions where the interface adapts ("adaptive interfaces") to each user and context dynamically.

We believe universal design and adaptive interfaces are a powerful combination for meeting the UX needs of people with disabilities, including SIIDs.

## UNIVERSAL DESIGN

It has been over 20 years since the Principles of Universal Design [3] were first introduced. The widespread success and adoption of universal design illustrates its value, especially in the physical built environment. However, the field of HCI is still learning the value of and how to apply universal design, partly due to the rapidly changing nature of digital information and experiences. Therefore, it is important to regularly identify how to adapt Universal Design Principles to HCI contexts.

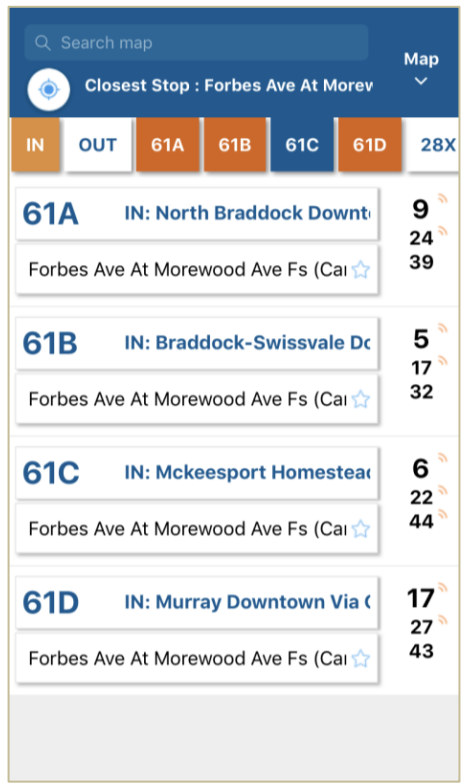

**Figure 1.** Tiramisu v3. Orange selections are pre-filters and blue are manual selections. Note that filtering results in all desired route rows to be seen above the fold.

Within the seven principles, several can be directly applied to the challenge of technology use during SIID. Examples, adapted from [7] for SIID include:

- **Flexible Use:** UIs should support use by either hand and support differences in user pace.
- **Perceptible Information:** Displays should have sufficient contrast for easy reading in the intended environment.
- **Tolerance for Error:** UX design should anticipate accidental or unintended actions.

The Principles were later complimented with the Goals of Universal Design [2,14]. As with the principles, a subset of the eight goals are also relevant to technology use during SIID. Adapted examples for SIID include:

- **Comfort:** Design interactions to be within limits of body function and perception.
- **Understanding:** Use intuitive, clear, and unambiguous information displays.
- **Personalization:** Support individual preferences.
- **Cultural Appropriateness:** UX design should reflect cultural values and local contexts.

## ADAPTIVE INTERFACES

The goal of Personalization is similar to the recent incorporation of machine learning into UX. Currently, the UX community is still investigating how to make adaptive interactions effective [15], but our research suggests adaptive interfaces have potential value for people with disabilities, including people with SIIDs. The potential is particularly pronounced for mobile, smartphone, and watch interactions due to smaller screen sizes and input choices.

Adaptive interfaces can also adapt to tasks and situations, rather than just user preferences. For example, the Virtual Information Officer adaptively extracted a variety of tasks from emails and the RADAR system guided task management for white collar workers [4,16]. In a mobile environment, a watch could alter the density of information when it detects user activities that would reduce visual perception ability. This adaption would likely be user-specific. For example, people will have differing skill with holding their wrist stable while walking.

## STRATEGY IN PRACTICE

As mentioned, we have research experience that suggests the combination of universal design and adaptive interfaces can be effective at supporting the needs of people with disabilities, including SIID. The Tiramisu Transit app was updated in version 3 to learn what information riders need based on what they previously needed at the same times and locations (Figure 1). The system can automatically pre-filter transit information, reducing the interaction effort required for people with visual, cognitive, or dexterity disabilities to obtain information. Instead of finding and selecting arrival information for desired routes from a long list of upcoming buses, the system only shows information inferred from the user's currently predicted needs, with options for correction, if necessary. Current inputs to the machine learning include a unique userID, time of day, location, and a history of past user interactions.

This UX is an example of universal design combined with adaptive interfaces in a way that supports SIID, since this has broad value to mobile users, especially when it is difficult or undesirable to expend more than a simple glance at the phone's screen. Our initial deployment with this feature suggests this approach is feasible and an A-B study is underway.

**RISKS**

One possible concern is the risk an induced disability acquired over time through learned dependency [1]. This disability may be as simple as deskilling or loss of mental models of operation (e.g., loss of map navigation skills [8]), but there is also the risk that extended learned dependencies could enter into the traditional definition of disability. Likewise, interfaces for SIID could exacerbate other problems stemming from digital experiences. Technology has already shown to influence the severity of depression and addiction in certain users [6]. Increasing access to information in normally digital-free situations could be counterproductive or harmful.

Similarly, there are signs that mobile technology use while walking has increased safety risk to pedestrians [9,11]. Research has already demonstrated reduced vehicle driving performance when responding to smartwatches [5], so it is logical to also assume interfaces for SIIDs could also increase distraction at inopportune times. Therefore, it is important that interfaces for SIID have the ability to either have contextual awareness or utilize low distraction approaches. For example, people who are blind often prefer bone conduction headphones when mobile since they do not physically obstruct ambient sounds.

**DISCUSSION**

There is great promise in creating effective technology experiences during SIID through the combination of universal design and adaptive interfaces. Research in each of these areas, and the combination, illustrates this promise.

Work is underway to assess the value of this combination in the real-world. Likewise, we expect additional studies on this topic in the near future. This body of work will likely quantify the value of this strategy and hopefully inspire new UX in products and research questions.

ACKNOWLEDGMENTS

The authors would like to thank Edward Steinfeld, Jordana Maisel, Qian Yang, and other members of our research team for their insights on universal design and adaptive interfaces. The contents of this paper were developed under grants from the National Institute on Disability, Independent Living, and Rehabilitation Research (NIDILRR grant numbers 90RE5011, 90DP0061, 90REGE0007). NIDILRR is a Center within the Administration for Community Living (ACL), Department of Health and Human Services (HHS).